\documentclass[aps,prl,twocolumn,showpacs,groupaddress]{revtex4}
\usepackage{amssymb}
\usepackage{graphicx}
\usepackage{dcolumn}
\usepackage{bm}
\usepackage{amsmath}

\begin{document}

\title{Comment on "Can Two-Photon Correlation of Chaotic Light be Considered as Correlation of Intensity Fluctuations ?"}
\author{Avi Pe'er}
\affiliation{JILA, National Institute of Standards and Technology
and Department of Physics, University of Colorado, Boulder, CO
80309-0440, USA}

\begin{abstract}

A Comment on the Letter by G. Scarcelli, V. Berardi and Y. Shih,
Phys. Rev. Lett. 96, 063602 (2006).
\end{abstract}

\maketitle

In a recent letter \cite{Scarcelli@Shih(bad)_PRL_2005}, Scarcelli,
Berardi and Shih discuss their recent experiments that
demonstrated ghost imaging by two-photon correlation with chaotic
light. The main claim of the letter is that the observed
two-photon correlation cannot be explained by a classical model.
Here I show how a classical reasoning similar to the classical
explanation of the Hanbury-Brown and Twiss experiment
\cite{Hanbury@Twiss_Nature_1956}, can fully account for the
results.\

In the experiment, a beam of chaotic light from a pseudo-thermal
source (a laser beam that it's spatial phase is randomized by a
moving diffuser) is split by a beam splitter; one beam is sent
through an aperturing object to a large bucket detector and the
other is sent to a small detector, positioned at the same distance
from the beam splitter as the object. The ghost image is detected by
measuring the intensity correlation between the two detectors as the
position of the small detector is scanned across the beam.\

Classically, chaotic thermal light can be considered as a random
noisy field $E(x,t)$ with a transverse correlation length (speckle
size) $l_c$ and correlation time $\tau_c$ that are short compared
to the dimensions of the beam and the integration time,
respectively. The beam splitter later produces two identical
copies of this noisy field $E_{1}(z=0,x,t)=E_{2}(z=0,x,t)$, which
propagate one to the object and the other to the small detector.
Since the two copies are identical they remain so during
propagation of any distance $z$, so the field (and intensity)
distributions are the same at any two planes of equal distance
from the beam splitter
\begin{eqnarray}
\label{Shih_eq1} E_{1}(z,x,t)=E_{2}(z,x,t). \
\end{eqnarray}
It is therefore clear that regardless of the distance from the
source or the beam splitter, the intensities at equivalent points
in the two beams are fully correlated, while at points that differ
by more than the speckle size they are uncorrelated. Assuming
Poissonian statistics for the spatial intensity distribution, the
correlation between equivalent points can be shown to be two times
larger than the uncorrelated background, as one expects for
thermal light \cite{Goodman_StatisticalOptics_1996}.\

Let us consider first a simple object that is composed of two
small holes in an opaque screen at points $y_{1},y_{2}$, each hole
is small compared to the correlation length of the intensity and
the distance between them is large compared to the correlation
length. It is clear that as the small detector is scanned across
the equivalent plane, cross correlation peaks will appear at
$x=y_{1}$ and $x=y_{2}$, exactly as shown in Fig. 2 of
\cite{Scarcelli@Shih(bad)_PRL_2005}. Since only one of the two
holes can contribute to the correlation at every point, the signal
to background ratio is only 3:2 in this measurement, and will
continue to degrade as the object becomes more complex, which is a
consequence of the use of a bucket detector. Yet, contrary to the
authors claim, the use of a bucket detector does not wash out the
correlation completely. In fact, for an object aperture of area
$A$ and small detector area $a$, the signal to background ratio
$S$ is
\begin{eqnarray}
\label{Shih_eq2} S=\frac{1+\left(A/a\right)}{\left(A/a\right)}, \
\end{eqnarray}
The area of the small detector is limited by the speckle area
$l_c^2$, which imposes the diffraction limit on the imaging
resolution. Consequently, a tradeoff exists between the imaging
resolution (small detector area $a$), and the detection contrast
($S$). For a given $a$, the maximal area of the object is limited
by the noise in the correlation measurement. \

The inclusion of a lens in one arm of the experiment images (and
magnifies) the intensity distribution from a plane $z=z_{1}$ to
another plane $z=z_{2}$, but does not affect the basic concept.
Accordingly, the intensity cross-correlation properties discussed
previously, are just copied (and stretched) from one plane to the
other, which explains the results shown in figure 3 of
\cite{Scarcelli@Shih(bad)_PRL_2005}.\

It is true, as the authors claim that the results cannot be
modelled by the Hanbury-Brown and Twiss formula, but this is only
because this formula inherently assumes far-field (it was
developed originally for astronomical measurements). Yet, the
basic phenomenon here and in the Hanbury-Brown Twiss experiment is
exactly the same, since the inherent correlation between the two
beams exists in any plane from near field to far field.\

\end{document}